\documentclass[aps,prl,twocolumn,amsfonts,amssymb,amsmath,floats,floatfix,showpacs,preprintnumbers,superscriptaddress]{revtex4-1}

\usepackage{graphicx}
\usepackage{hyperref}
\pdfoutput=1 
\hypersetup{
colorlinks=true, 
urlcolor= blue, 
citecolor=blue,
linkcolor= blue, 
bookmarks=true, 
bookmarksopen=false, 
    pdfauthor   = {Giorgio Levy},
    pdftitle    = {RPES on Iron-pnictides},
    pdfkeywords = {RPES on Iron-pnictides}, }
\usepackage[pdftex]{thumbpdf}       
\begin{document}

\title{Probing the role of Co substitution in the electronic structure of iron-pnictides}

\author{G. Levy}
\email{levyg@physics.ubc.ca}
\affiliation{Department of Physics {\rm {\&}} Astronomy, University of British Columbia, Vancouver, British Columbia V6T\,1Z1, Canada}
\author{R. Sutarto}
\affiliation{Department of Physics {\rm {\&}} Astronomy, University of British Columbia, Vancouver, British Columbia V6T\,1Z1, Canada}
\affiliation{Canadian Light Source, University of Saskatchewan, Saskatoon, Saskatchewan S7N 0X4, Canada}
\author{D. Chevrier}
\author{T. Regier}
\author{R. Blyth}
\affiliation{Canadian Light Source, University of Saskatchewan, Saskatoon, Saskatchewan S7N 0X4, Canada}
\author{J. Geck}
\author{S. Wurmehl}
\author{L.\,Harnagea} 
\affiliation{Leibniz Institute for Solid State and Materials Research IFW Dresden, 01069 Dresden, Germany}
\author{H. Wadati}
\affiliation{Department\,of\,Applied\,Physics\,and\,Quantum-Phase\,Electronics\,Center,\,University\,of\,Tokyo,\,Hongo,\,Tokyo\,113-8656,\,Japan}
\author{T. Mizokawa}
\affiliation{Department\,of\,Applied\,Physics\,and\,Quantum-Phase\,Electronics\,Center,\,University\,of\,Tokyo,\,Hongo,\,Tokyo\,113-8656,\,Japan}
\author{I.S. Elfimov}
\author{A. Damascelli}
\author{G.A. Sawatzky}
\affiliation{Department of Physics {\rm {\&}} Astronomy, University of British Columbia, Vancouver, British Columbia V6T\,1Z1, Canada}
\affiliation{Quantum Matter Institute, University of British Columbia, Vancouver, British Columbia V6T\,1Z4, Canada}

\date{\today}

\begin{abstract}
The role of Co substitution in the low-energy electronic structure of Ca(Fe$_{0.944}$Co$_{0.056}$)$_2$As$_2$ is investigated by resonant photoemission spectroscopy and density functional theory. The Co 3$d$-state center-of-mass is observed at 250\,meV higher binding energy than Fe's, indicating that Co posses one extra valence electron, and that Fe and Co are in the same 2+ oxidation state. Yet, significant Co character is detected for the Bloch wavefunctions at the chemical potential, revealing that the Co 3$d$ electrons are part of the Fermi sea determining the Fermi surface. This establishes the complex role of Co substitution in CaFe$_2$As$_2$, and the inadequacy of a rigid-band shift description.
\end{abstract}


\maketitle

The physical properties of iron-arsenide compounds can be tuned by substituting Fe with Co. For instance, in regard to CaFe$_2$As$_2$ and its $\sim\!170$\,K transition from paramagnetic tetragonal to antiferromagnetic orthorhombic phase \cite{ronning2008, ni2008}, the characteristic temperature is rapidly suppressed when Fe is replaced by Co -- disappearing for Co concentrations around 7\% \cite{Harnagea2010}. For even higher Co substitution, a superconducting phase emerges with a maximum critical temperature of about 20\,K.  

The specific mechanism via which transition-metal substitution leads to these effects is still highly debated, with carrier density variation and impurity scattering -- as well as their intimate interplay -- being the main scenarios under consideration \citep{wadati2010, neupane2011, Berlijn2011, Haverkort2011, Liu2012}. The proposal that Co might donate one electron to the system, effectively doping it as in a rigid chemical-potential shift within an unperturbed band structure, is qualitatively supported by angle-resolved photo\-emission spectroscopy studies \cite{liu2010}, which provide evidence for the disappearance of the hole pocket at the Brillouin zone center upon Co substitution. Alternatively, it has been argued that Co is isovalent to Fe, and that the main role of the Fe-Co substitution is to introduce a random impurity potential \citep{wadati2010}. This would lead to scattering of the itinerant charge carriers, consistent with a non-vanishing imaginary part of the self-energy even at the Fermi level, as proposed in Ref. \onlinecite{wadati2010} and demonstrated in more recent calculations \citep{Berlijn2011, Haverkort2011}. In turn, possible nesting vectors connected to the onset of magnetic ordering are smeared out, and with it also the Fermi surface and its direct relation to Luttinger's counting.

\begin{figure*}[htb]
\includegraphics{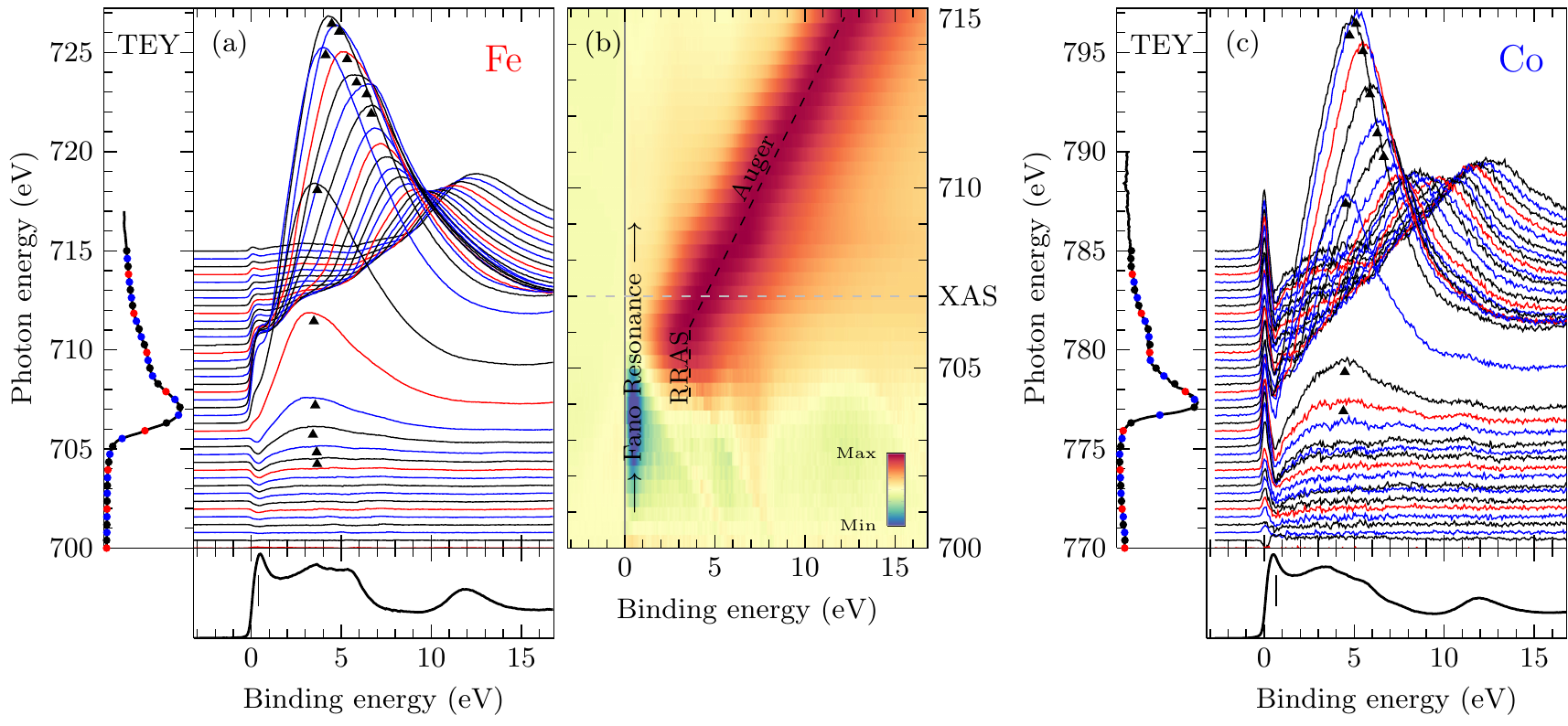}\vspace{-.6cm}
 \caption{\label{respes} (color online) Evolution of the Ca(Fe$_{0.944}$Co$_{0.056}$)$_2$As$_2$ resonant photoemission spectra (RPES) upon scanning the photon energy across the (a,b) Fe and (c) Co $L_3$ absorption edges; the RPES curves, taken in 0.395 eV steps, have been displaced vertically to match the energy scale of the Fe and Co edges shown on the left-hand side of (a,c). The non-resonant PES signal shown at the bottom of (a,c), averaged over 3 spectra at 700 (770) eV for Fe (Co), was subtracted to highlight the RPES behavior. In (b) a false color plot of the spectra from (a), normalized to their total area, is shown; the different characteristics of RPES are emphasized: radiationless Raman Auger scattering (RRAS), Auger emission, and Fano resonance.}
\end{figure*}

To experimentally determine the role of Co-induced states, we study Ca(Fe$_{0.944}$Co$_{0.056}$)$_2$As$_2$ by resonant photoemission spectroscopy (RPES), which provides the advantage of element selectivity through the involvement of a core-electron excitation in the resonant photon-absorption process. We show that the center of mass of the Co-induced low-energy states is at 250 meV higher binding energy than Fe's, which provides a direct measure of the Co impurity potential. The screening of the latter, as revealed by the experimental estimate of $U_{dd}$ for Fe and Co and a density functional theory (DFT) analysis, leads to 1 extra 3$d$-electron being associated with Co and, in turn, to the isovalence of Fe and Co. Yet, the Bloch states near the chemical potential have significant Co character. These findings point to the inadequacy of the rigid-band scenario and to the more active role of Co in determining the properties on these materials.

The RPES experiments were performed at the Canadian Light Source SGM Beamline on the (001) surface of Ca(Fe$_{0.944}$Co$_{0.056}$)$_2$As$_2$. The single crystals were grown from Sn flux \cite{Harnagea2010}, and the $x=0.056$ Co concentration was determined by energy dispersive x-ray spectroscopy. The samples were cleaved in situ and maintained at 300\,K and pressures better than $2\!\times\!10^{-8}$ Torr. Absorption spectra were acquired by the total electron yield (TEY) technique, normalized to the beam flux; RPES spectra were measured with horizontal polarization, a Scienta\,100 hemispherical analyzer, and $\sim 0.15$ eV energy resolution as calibrated on a Au film. 

In RPES, the valence-band photoemission signal is measured while varying the photon energy across an elemental x-ray absorption edge; in addition to photoemitting a valence electron, the photon can also excite an electron from the resonating core level into an empty state just above the chemical potential. The subsequent nonradiative decay of the core hole through various Auger electron emission channels, and the interference between direct photoemission and Auger processes, lead to an element-specific enhancement and evolution of the photoemission signal \citep{gelmukhanov199987, bruhwiler2002}.  Here we will investigate the RPES for photon energy resonating with the $L_3$ absorption edge of Fe (707\,eV) and Co (777\,eV), as shown in Fig.\,\ref{respes}(a) and (c) respectively, corresponding to the transition $2p^6~3d^n \rightarrow 2p^5~3d^{n+1}$. In Fig.\,\ref{respes}(b) we present a false color plot of normalized Fe-edge data, highlighting the typical RPES spectral features \citep{gelmukhanov199987, bruhwiler2002}: (i) a photoemission enhancement -- at constant binding energy -- due to radiationless Raman Auger scattering (RRAS), when the photoexcited core-electron acts as a {\it spectator} to the core-hole recombination process; (ii) a photoemission peak  -- evolving linearly with photon energy -- due to conventional Auger emission, when the photoexcited electron from the core {\it delocalizes} faster than the core-hole  lifetime; (iii) a Fano resonance due to quantum interference observed near the chemical potential as a function of photon energy \citep{fano1961}, when the photoexcited core-electron acts as a {\it participant} in the core-hole recombination.

\begin{figure}[t!]
\includegraphics{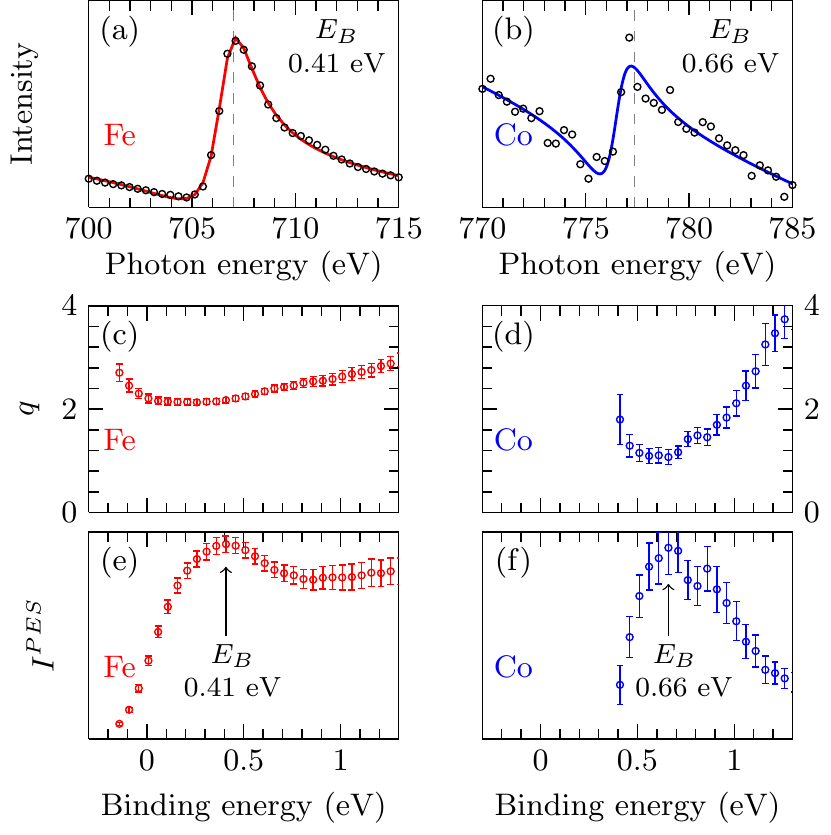}
\caption{\label{characteristics} (color online). (a,b) Fano profile observed by cutting the RPES data from Fig.\,\ref{respes}(a,c) as a function of photon energy and at fixed binding energy: $E_B\!=\!0.41$ and 0.66\,eV for Fe and Co, respectively (vertical dashed lines mark the energy of maximum x-ray absorption, 707.1 and 777.3\,eV). Red and blue lines are a fit to Eq.\,\ref{eq:fanoline}; the corresponding Fano asymmetry parameter $q$ and non-resonant photoemission intensity $I^{PES}$ are shown in panel (c,d) and (e,f), for a fit performed over the binding energy range for which a Fano profile with $q\!<\!4$ is observed (the smaller the $q$, the larger the asymmetry).}
\end{figure}

\begin{figure}
\includegraphics{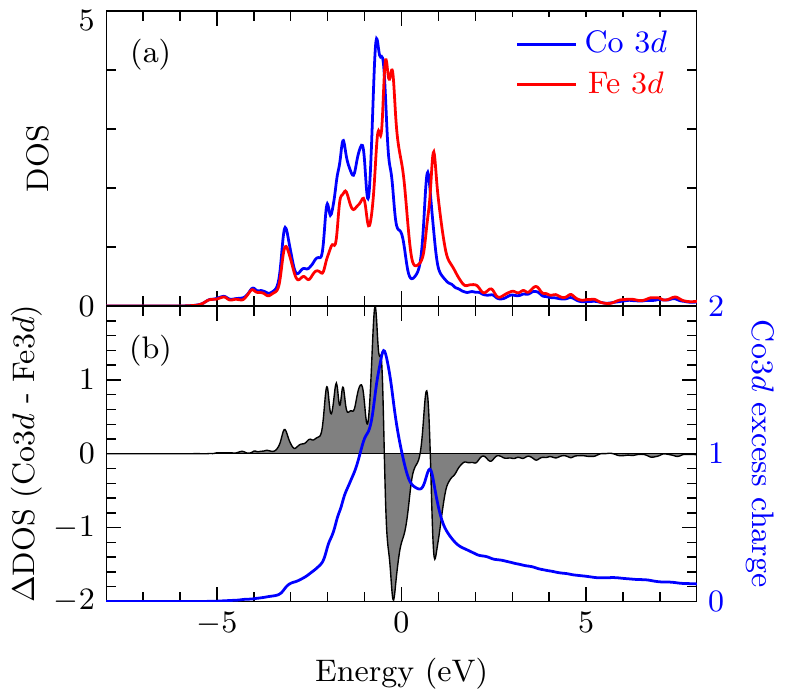}
\caption{\label{DFT} (color online). (a) Partial Fe (red) and Co (blue) $3d$-density of states (DOS) from supercell calculations for pure and  5.6\% Co-substituted CaFe$_2$As$_2$, respectively; the center of gravity for Co is $\sim$0.25 eV deeper in binding energy than for Fe. (b) Difference between Co and Fe DOS (filled area, left axis), and its corresponding integral (blue line, right axis); the latter, when estimated at the Fermi level ($E_F=0$), is a measure of the excess charge associated with Co: $\sim$1 electron.}
\end{figure}

In studying the role of Co substitution by PES, the main shortcoming is that non-resonant PES does not allow to identify the Co contribution to the low-energy electronic structure. The Ca(Fe$_{0.944}$Co$_{0.056}$)$_2$As$_2$ non-resonant valence-band spectra shown in the bottom panels of Fig.\,\ref{respes}(a) and (c) -- measured at $\sim\!700$ and 770\,eV photon energy respectively, thus away from the corresponding $L_3$ absorption edges (left panels) -- are very similar to the results reported for Co-free CaFe$_2$As$_2$ \citep{Jong2009}. The detected structure can be associated with the one-electron removal from Fe-$3d$ (0-2\,eV), As-$4p$ (3-6\,eV), and As-$4s$ (11-13\,eV), with no additional identifiable peak stemming from the presence of Co \citep{Jong2009}. This is also consistent with DFT calculations \cite{wadati2010,Arita1,Arita2}, which predict a very similar 3$d$ partial density of states (DOS) for Fe and Co close to the chemical potential, with only a relative energy shift (Fig.\,\ref{DFT}). This shortcoming can be overcome by taking advantage of the element specificity of RPES.

The RPES spectra obtained by varying the photon energy across the $L_3$ absorption edge of either Fe and Co are presented in Fig.\,\ref{respes}(a,b) and (c) [after subtraction of the non-resonant spectra shown in the bottom panels of (a) and (c), to emphasize the resonant behavior]. The most obvious features -- see triangles in (a,c) and red area in (b)-- are the signal enhancement associated with the constant binding-energy RRAS and, upon increasing the photon energy, the constant kinetic-energy conventional Auger emission (this appears as linearly dispersing when plotted vs. binding energy). The transition between the two regimes as a function of photon energy is defined by the crossing between the constant and linearly dispersive behavior, as shown for the case of Fe by the two dashed lines in Fig.\,\ref{respes}(b), which takes place at a photon energy of 0.9\,eV (0.5) below the absorption maximum located at 707.1\,eV (777.3) for Fe (Co). As for the observed RRAS binding energy value for Fe (3.6\,eV) and Co (4.5\,eV) it should be noted that, since the RRAS can be thought of as a {\it two-hole}$-${\it one-electron} state, the 0.9\,eV Fe-Co offset stems from the difference in on-site Coulomb repulsion $U_{dd}$ for the two elements. Indeed, this 0.9\,eV offset is in good agreement with the difference between the Cini-Sawatzky  \citep{M1976605, PhysRevLett.39.504} estimate of $U_{dd}^{Fe}\!=\!1.50$\,eV and $U_{dd}^{Co}\!=\!2.75$\,eV from combined XPS and Auger spectroscopy (not shown). Most important, this approximately  1\,eV difference is consistent with what is found in other metallic systems of Co and Fe isovalent impurities \citep{PhysRevB.37.10674}, suggesting  that in Co-substituted iron-pnictides Fe and Co are in the same oxidation state.

An interesting aspect of the RPES data in Fig.\,\ref{respes} is found at lower binding energies, where an asymmetric Fano profile \citep{fano1961} is detected when cutting the RPES 2-dimensional dataset at a fixed binding energy and plotting the RPES signal as a function of photon energy, as shown in Fig.\,\ref{characteristics}(a,b) for Fe and Co. When the photoexcited core-electron participates in the Auger decay of the core-hole, the final state is the same as the one reached in direct photoemission from the valence band; the interference between these two parallel channels leading to the same final state, and their overlapping discrete (Auger) and continuum (PES) character in energy, produces a characteristic Fano lineshape \citep{fano1961}. After subtraction of a linear background, this can be written as:
\begin{equation}
 I^{RPES}(\hbar \omega) = I^{PES}(\hbar \omega) \frac{\,(q+E)^2}{1+E^2},  
\label{eq:fanoline}
\end{equation}
where $\hbar \omega$  is the incident photon energy, $I^{PES}$ is the PES intensity in the direct channel, $E\!=\!(\hbar \omega\!-\!E_R)/2\Gamma_R$ with $E_R$ and $\Gamma_R$ being the resonance energy and half width, and $q$ is the dimensionless Fano asymmetry parameter (a Lorentzian lineshape is recovered for $|q|\!\rightarrow\!\infty$).

We use Eq.\,\ref{eq:fanoline} plus a linear background to fit the RPES spectra as a function of photon energy, in a $\sim$2\,eV binding energy range about the chemical potential. From the $q$ parameter values presented in Fig.\,\ref{characteristics}(c,d) for Fe and Co, we observe that the most pronounced asymmetries -- corresponding to the largest interference effects -- are found in slightly different binding energy regions for Fe (from 0 to -0.4\,eV) than for Co (from -0.5 to -0.9\,eV); at the same time, the value of $I^{PES}$ is maximum at $E_B\!=\!0.41$\,eV (Fe) and 0.66\,eV (Co), as shown in  Fig.\,\ref{characteristics}(e,f). These binding energies, which are also indicated by vertical bars in the lower panels of Fig.\,\ref{respes}(a,c) and whose corresponding Fano profiles are shown in Fig.\,\ref{characteristics}(a,b), identify the the characteristic average energy for Fe and Co states in Co-substituted CaFe$_2$As$_2$. The observed 0.25\,eV larger energy for the single-electron removal from Co provides a direct measure of the Co impurity potential.

This $\Delta E_B\!\simeq\!0.25$\,eV shift between Fe and Co electron-removal energies is consistent with ab-initio DFT calculations \cite{wadati2010,Arita1,Arita2}. For the most accurate quantitative comparison with our experimental results from Ca(Fe$_{0.944}$Co$_{0.056}$)$_2$As$_2$, we have performed a supercell calculation using the Wien2K package, with one out of 18 Fe atoms replaced by a Co atom, corresponding to a Co   concentration of 5.6\%. The unit cell parameters for the high-temperature tetragonal phase of CaFe$_2$As$_2$ are obtained from Kreyssig {\it et al.} \citep{Kreyssig2008}, as determined by Rietveld analysis of neutron diffraction experiments: $a\equiv b\!=\!(3.912\pm 0.068)$\AA, $c\!=\!(11.667 \pm 0.045)$\,\AA, and $z_{As}\!=\!0.35814$\,\AA. The resulting partial Co 3$d$-DOS is shown in Fig.\,\ref{DFT}(a), together with the one of Fe calculated for Co-free CaFe$_2$As$_2$ with the same approach. This comparison reveals a relative 0.25\,eV shift for the center of mass of Fe and Co 3$d$-DOS (as calculated from the difference in first moments in the range $-\!8$ to 0\,eV), in agreement with the $\Delta E_B$ determined experimentally from the Fano resonance analysis of RPES. The impurity potential associated with this shift leads to a screening charge accumulation around Co, which can be estimated from the difference between Fe and Co 3$d$-DOS [shaded are in Fig.\,\ref{DFT}(b), left axis]. In particular, by integrating in energy $\Delta$DOS(Co$3d-$Fe$3d$) from the bottom of the 3$d$ DOS to the chemical potential [blue line in Fig.\,\ref{DFT}(b), right axis], one obtains that Co is surrounded by 1 extra 3$d$-electron as compared to Fe (note that the difference vanishes when the integration is performed over the full 3$d$-DOS including the unoccupied states, as expected since the total number of 3$d$ states is the same). This again points to the isovalence of Co and Fe in this compound -- consistent with the $U_{dd}$ analysis -- and more specifically to a 2+ oxidation state for both Co and Fe.

\begin{figure}
\includegraphics{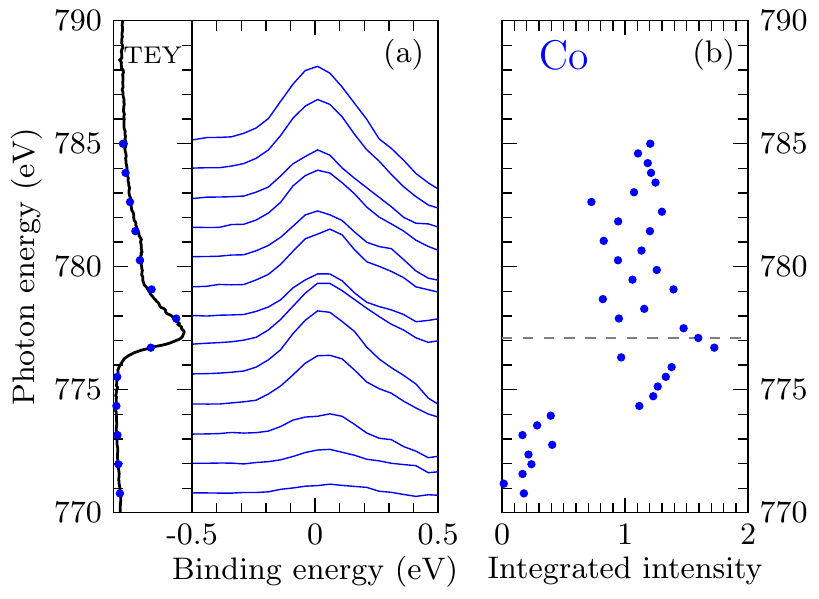}
\caption{\label{hyb}  Evolution of the Co RPES spectra near $E_F$ upon scanning the photon energy; the curves have been displaced vertically to match the energy scale of the Co absorption edge shown in the left-hand side panel. (b) Integrated intensity of the near-$E_F$ peak vs. photon energy, showing a cross-section enhancement starting at the Co $L_3$ edge (dashed line).}
\end{figure}

The close inspection of the Co-edge RPES spectra at the chemical potential provides a last additional clue on the role of Co in the low-energy electronic structure of Co-substituted CaFe$_2$As$_2$. As shown in Fig.\,\ref{respes}(c), and in greater detail in Fig.\,\ref{hyb}, in the binding energy range from 0.5 to $-\!0.5$\,eV, where a Fano lineshape was not detected [$I^{PES}$ is vanishing below $-\!0.4$\,eV in Fig.\,\ref{characteristics}(f)], one can observe a peak whose intensity exhibits a somewhat resonating behavior. To track its photon energy evolution, we fit this feature with a Voigt function with a Gaussian width of 0.15\,eV to account for the energy resolution; the integrated intensity enhancement at photon energies larger than 775\,eV [i.e. the leading edge of the Co $L_3$ absorption, Fig.\,\ref{hyb}(b)] clearly demonstrate the presence of Co character also at the chemical potential. This indicates that -- despite its impurity nature -- Co also contributes 3$d$ states to the Bloch wavefunctions at $E_F$, and thus to the details of Fermi surface and Fermi sea.

In conclusion, by taking advantage of the element specificity of RPES we probed the role of Co substitution in CaFe$_2$As$_2$. The observed $\Delta E_B\!\simeq\!0.25$\,eV shift between Co and Fe single-electron excitations, consistent with ab-initio DFT calculations and estimates of the on-site Coulomb repulsion $U_{dd}$, provides a quantitative determination of the Co impurity potential and the 2+ isovalence of Co and Fe. This, together with the detection of a Co participation in the electronic states belonging to the Fermi surface, establishes the complex role of Co substitution beyond a mere rigid-band shift description.

This work was supported by the Killam, Sloan, CRC (A.D and G.A.S.), and NSERC's Steacie Fellowship Programs (A.D.), JSPS and CSTP (H.W.), NSERC, CFI, CIFAR Quantum Materials, and BCSI. S.W. and J.G. acknowledge support by DFG under the Emmy-Noether program (Grant No. WU595/3-1 and GE1647/2-1), and S.W. under the Priority Programme SPP1458 (Grants No. BE1749/13). CLS is supported by NSERC, NRC, CIHR, and the University of Saskatchewan.

\bibliography{Levy_Co_FeAs_PRL}

\begin{thebibliography}{19}%
\makeatletter
\providecommand \@ifxundefined [1]{%
 \@ifx{#1\undefined}
}%
\providecommand \@ifnum [1]{%
 \ifnum #1\expandafter \@firstoftwo
 \else \expandafter \@secondoftwo
 \fi
}%
\providecommand \@ifx [1]{%
 \ifx #1\expandafter \@firstoftwo
 \else \expandafter \@secondoftwo
 \fi
}%
\providecommand \natexlab [1]{#1}%
\providecommand \enquote  [1]{``#1''}%
\providecommand \bibnamefont  [1]{#1}%
\providecommand \bibfnamefont [1]{#1}%
\providecommand \citenamefont [1]{#1}%
\providecommand \href@noop [0]{\@secondoftwo}%
\providecommand \href [0]{\begingroup \@sanitize@url \@href}%
\providecommand \@href[1]{\@@startlink{#1}\@@href}%
\providecommand \@@href[1]{\endgroup#1\@@endlink}%
\providecommand \@sanitize@url [0]{\catcode `\\12\catcode `\$12\catcode
  `\&12\catcode `\#12\catcode `\^12\catcode `\_12\catcode `\%12\relax}%
\providecommand \@@startlink[1]{}%
\providecommand \@@endlink[0]{}%
\providecommand \url  [0]{\begingroup\@sanitize@url \@url }%
\providecommand \@url [1]{\endgroup\@href {#1}{\urlprefix }}%
\providecommand \urlprefix  [0]{URL }%
\providecommand \Eprint [0]{\href }%
\providecommand \doibase [0]{http://dx.doi.org/}%
\providecommand \selectlanguage [0]{\@gobble}%
\providecommand \bibinfo  [0]{\@secondoftwo}%
\providecommand \bibfield  [0]{\@secondoftwo}%
\providecommand \translation [1]{[#1]}%
\providecommand \BibitemOpen [0]{}%
\providecommand \bibitemStop [0]{}%
\providecommand \bibitemNoStop [0]{.\EOS\space}%
\providecommand \EOS [0]{\spacefactor3000\relax}%
\providecommand \BibitemShut  [1]{\csname bibitem#1\endcsname}%
\let\auto@bib@innerbib\@empty
\bibitem [{\citenamefont {Ronning}\ \emph {et~al.}(2008)\citenamefont {Ronning}
  \emph {et~al.}}]{ronning2008}%
  \BibitemOpen
  \bibfield  {author} {\bibinfo {author} {\bibfnamefont {F.}~\bibnamefont
  {Ronning}} \emph {et~al.},\ }\href
  {http://stacks.iop.org/0953-8984/20/i=32/a=322201} {\bibfield  {journal}
  {\bibinfo  {journal} {J. Phys.: Cond. Matter}\ }\textbf {\bibinfo {volume}
  {20}},\ \bibinfo {pages} {322201} (\bibinfo {year} {2008})}\BibitemShut
  {NoStop}%
\bibitem [{\citenamefont {Ni}\ \emph {et~al.}(2008)\citenamefont {Ni} \emph
  {et~al.}}]{ni2008}%
  \BibitemOpen
  \bibfield  {author} {\bibinfo {author} {\bibfnamefont {N.}~\bibnamefont {Ni}}
  \emph {et~al.},\ }\href {\doibase 10.1103/PhysRevB.78.014523} {\bibfield
  {journal} {\bibinfo  {journal} {Phys. Rev. B}\ }\textbf {\bibinfo {volume}
  {78}},\ \bibinfo {pages} {014523} (\bibinfo {year} {2008})}\BibitemShut
  {NoStop}%
\bibitem [{\citenamefont {Harnagea}\ \emph {et~al.}(2011)\citenamefont
  {Harnagea} \emph {et~al.}}]{Harnagea2010}%
  \BibitemOpen
  \bibfield  {author} {\bibinfo {author} {\bibfnamefont {L.}~\bibnamefont
  {Harnagea}} \emph {et~al.},\ }\href {\doibase 10.1103/PhysRevB.83.094523}
  {\bibfield  {journal} {\bibinfo  {journal} {Phys. Rev. B}\ }\textbf {\bibinfo
  {volume} {83}},\ \bibinfo {pages} {094523} (\bibinfo {year}
  {2011})}\BibitemShut {NoStop}%
\bibitem [{\citenamefont {Wadati}\ \emph {et~al.}(2010)\citenamefont {Wadati},
  \citenamefont {Elfimov},\ and\ \citenamefont {Sawatzky}}]{wadati2010}%
  \BibitemOpen
  \bibfield  {author} {\bibinfo {author} {\bibfnamefont {H.}~\bibnamefont
  {Wadati}}, \bibinfo {author} {\bibfnamefont {I.}~\bibnamefont {Elfimov}}, \
  and\ \bibinfo {author} {\bibfnamefont {G.~A.}\ \bibnamefont {Sawatzky}},\
  }\href {\doibase 10.1103/PhysRevLett.105.157004} {\bibfield  {journal}
  {\bibinfo  {journal} {Phys. Rev. Lett.}\ }\textbf {\bibinfo {volume} {105}},\
  \bibinfo {pages} {157004} (\bibinfo {year} {2010})}\BibitemShut {NoStop}%
\bibitem [{\citenamefont {Neupane}\ \emph {et~al.}(2011)\citenamefont {Neupane}
  \emph {et~al.}}]{neupane2011}%
  \BibitemOpen
  \bibfield  {author} {\bibinfo {author} {\bibfnamefont {M.}~\bibnamefont
  {Neupane}} \emph {et~al.},\ }\href {\doibase 10.1103/PhysRevB.83.094522}
  {\bibfield  {journal} {\bibinfo  {journal} {Phys. Rev. B}\ }\textbf {\bibinfo
  {volume} {83}},\ \bibinfo {pages} {094522} (\bibinfo {year}
  {2011})}\BibitemShut {NoStop}%
\bibitem [{\citenamefont {Berlijn}\ \emph {et~al.}(2011)\citenamefont {Berlijn}
  \emph {et~al.}}]{Berlijn2011}%
  \BibitemOpen
  \bibfield  {author} {\bibinfo {author} {\bibfnamefont {T.}~\bibnamefont
  {Berlijn}} \emph {et~al.},\ }\href {http://arxiv.org/abs/1112.4858} {}
  (\bibinfo {year} {2011}),\ \Eprint {http://arxiv.org/abs/1112.4858}
  {arXiv:1112.4858} \BibitemShut {NoStop}%
\bibitem [{\citenamefont {Haverkort}\ \emph {et~al.}(2011)\citenamefont
  {Haverkort}, \citenamefont {Elfimov},\ and\ \citenamefont
  {Sawatzky}}]{Haverkort2011}%
  \BibitemOpen
  \bibfield  {author} {\bibinfo {author} {\bibfnamefont {M.~W.}\ \bibnamefont
  {Haverkort}}, \bibinfo {author} {\bibfnamefont {I.~S.}\ \bibnamefont
  {Elfimov}}, \ and\ \bibinfo {author} {\bibfnamefont {G.~A.}\ \bibnamefont
  {Sawatzky}},\ }\href {http://arxiv.org/abs/1109.4036} {} (\bibinfo {year}
  {2011}),\ \Eprint {http://arxiv.org/abs/1109.4036} {arXiv:1109.4036}
  \BibitemShut {NoStop}%
\bibitem [{\citenamefont {Liu}\ and\ \citenamefont {Zhou}(2012)}]{Liu2012}%
  \BibitemOpen
  \bibfield  {author} {\bibinfo {author} {\bibfnamefont {S.~L.}\ \bibnamefont
  {Liu}}\ and\ \bibinfo {author} {\bibfnamefont {T.}~\bibnamefont {Zhou}},\
  }\href {http://arxiv.org/abs/1201.0363} {} (\bibinfo {year} {2012}),\ \Eprint
  {http://arxiv.org/abs/1201.0363} {arXiv:1201.0363} \BibitemShut {NoStop}%
\bibitem [{\citenamefont {Liu}\ \emph {et~al.}(2010)\citenamefont {Liu} \emph
  {et~al.}}]{liu2010}%
  \BibitemOpen
  \bibfield  {author} {\bibinfo {author} {\bibfnamefont {C.}~\bibnamefont
  {Liu}} \emph {et~al.},\ }\href {\doibase 10.1038/nphys1656} {\bibfield
  {journal} {\bibinfo  {journal} {Nat. Phys.}\ }\textbf {\bibinfo {volume}
  {6}},\ \bibinfo {pages} {419} (\bibinfo {year} {2010})}\BibitemShut {NoStop}%
\bibitem [{\citenamefont {Gel'mukhanov}\ and\ \citenamefont
  {$\AA$gren}(1999)}]{gelmukhanov199987}%
  \BibitemOpen
  \bibfield  {author} {\bibinfo {author} {\bibfnamefont {F.}~\bibnamefont
  {Gel'mukhanov}}\ and\ \bibinfo {author} {\bibfnamefont {H.}~\bibnamefont
  {$\AA$gren}},\ }\href {\doibase 10.1016/S0370-1573(99)00003-4} {\bibfield
  {journal} {\bibinfo  {journal} {Phys. Rep.}\ }\textbf {\bibinfo {volume}
  {312}},\ \bibinfo {pages} {87 } (\bibinfo {year} {1999})}\BibitemShut
  {NoStop}%
\bibitem [{\citenamefont {Br\"uhwiler}\ \emph {et~al.}(2002)\citenamefont
  {Br\"uhwiler}, \citenamefont {Karis},\ and\ \citenamefont
  {M\aa{}rtensson}}]{bruhwiler2002}%
  \BibitemOpen
  \bibfield  {author} {\bibinfo {author} {\bibfnamefont {P.~A.}\ \bibnamefont
  {Br\"uhwiler}}, \bibinfo {author} {\bibfnamefont {O.}~\bibnamefont {Karis}},
  \ and\ \bibinfo {author} {\bibfnamefont {N.}~\bibnamefont {M\aa{}rtensson}},\
  }\href {\doibase 10.1103/RevModPhys.74.703} {\bibfield  {journal} {\bibinfo
  {journal} {Rev. Mod. Phys.}\ }\textbf {\bibinfo {volume} {74}},\ \bibinfo
  {pages} {703} (\bibinfo {year} {2002})}\BibitemShut {NoStop}%
\bibitem [{\citenamefont {Fano}(1961)}]{fano1961}%
  \BibitemOpen
  \bibfield  {author} {\bibinfo {author} {\bibfnamefont {U.}~\bibnamefont
  {Fano}},\ }\href {\doibase 10.1103/PhysRev.124.1866} {\bibfield  {journal}
  {\bibinfo  {journal} {Phys. Rev.}\ }\textbf {\bibinfo {volume} {124}},\
  \bibinfo {pages} {1866} (\bibinfo {year} {1961})}\BibitemShut {NoStop}%
\bibitem [{\citenamefont {de~Jong}\ \emph {et~al.}(2009)\citenamefont {de~Jong}
  \emph {et~al.}}]{Jong2009}%
  \BibitemOpen
  \bibfield  {author} {\bibinfo {author} {\bibfnamefont {S.}~\bibnamefont
  {de~Jong}} \emph {et~al.},\ }\href {\doibase 10.1103/PhysRevB.79.115125}
  {\bibfield  {journal} {\bibinfo  {journal} {Phys. Rev. B}\ }\textbf {\bibinfo
  {volume} {79}},\ \bibinfo {pages} {115125} (\bibinfo {year}
  {2009})}\BibitemShut {NoStop}%
\bibitem [{\citenamefont {Nakamura}\ \emph {et~al.}(2011)\citenamefont
  {Nakamura}, \citenamefont {Arita},\ and\ \citenamefont {Ikeda}}]{Arita1}%
  \BibitemOpen
  \bibfield  {author} {\bibinfo {author} {\bibfnamefont {K.}~\bibnamefont
  {Nakamura}}, \bibinfo {author} {\bibfnamefont {R.}~\bibnamefont {Arita}}, \
  and\ \bibinfo {author} {\bibfnamefont {H.}~\bibnamefont {Ikeda}},\
  }\href@noop {} {\bibfield  {journal} {\bibinfo  {journal} {Phys. Rev. B}\
  }\textbf {\bibinfo {volume} {83}},\ \bibinfo {pages} {144512} (\bibinfo
  {year} {2011})}\BibitemShut {NoStop}%
\bibitem [{\citenamefont {Konbu}\ \emph {et~al.}(2011)\citenamefont {Konbu}
  \emph {et~al.}}]{Arita2}%
  \BibitemOpen
  \bibfield  {author} {\bibinfo {author} {\bibfnamefont {S.}~\bibnamefont
  {Konbu}} \emph {et~al.},\ }\href@noop {} {\bibfield  {journal} {\bibinfo
  {journal} {J. Phys. Soc. Jpn.}\ }\textbf {\bibinfo {volume} {80}},\ \bibinfo
  {pages} {123701} (\bibinfo {year} {2011})}\BibitemShut {NoStop}%
\bibitem [{\citenamefont {Cini}(1976)}]{M1976605}%
  \BibitemOpen
  \bibfield  {author} {\bibinfo {author} {\bibfnamefont {M.}~\bibnamefont
  {Cini}},\ }\href {\doibase 10.1016/0038-1098(76)91070-X} {\bibfield
  {journal} {\bibinfo  {journal} {Solid State Comm.}\ }\textbf {\bibinfo
  {volume} {20}},\ \bibinfo {pages} {605 } (\bibinfo {year}
  {1976})}\BibitemShut {NoStop}%
\bibitem [{\citenamefont {Sawatzky}(1977)}]{PhysRevLett.39.504}%
  \BibitemOpen
  \bibfield  {author} {\bibinfo {author} {\bibfnamefont {G.~A.}\ \bibnamefont
  {Sawatzky}},\ }\href {\doibase 10.1103/PhysRevLett.39.504} {\bibfield
  {journal} {\bibinfo  {journal} {Phys. Rev. Lett.}\ }\textbf {\bibinfo
  {volume} {39}},\ \bibinfo {pages} {504} (\bibinfo {year} {1977})}\BibitemShut
  {NoStop}%
\bibitem [{\citenamefont {van~der Marel}\ and\ \citenamefont
  {Sawatzky}(1988)}]{PhysRevB.37.10674}%
  \BibitemOpen
  \bibfield  {author} {\bibinfo {author} {\bibfnamefont {D.}~\bibnamefont
  {van~der Marel}}\ and\ \bibinfo {author} {\bibfnamefont {G.~A.}\ \bibnamefont
  {Sawatzky}},\ }\href {\doibase 10.1103/PhysRevB.37.10674} {\bibfield
  {journal} {\bibinfo  {journal} {Phys. Rev. B}\ }\textbf {\bibinfo {volume}
  {37}},\ \bibinfo {pages} {10674} (\bibinfo {year} {1988})}\BibitemShut
  {NoStop}%
\bibitem [{\citenamefont {Kreyssig}\ \emph {et~al.}(2008)\citenamefont
  {Kreyssig} \emph {et~al.}}]{Kreyssig2008}%
  \BibitemOpen
  \bibfield  {author} {\bibinfo {author} {\bibfnamefont {A.}~\bibnamefont
  {Kreyssig}} \emph {et~al.},\ }\href {\doibase 10.1103/PhysRevB.78.184517}
  {\bibfield  {journal} {\bibinfo  {journal} {Phys. Rev. B}\ }\textbf {\bibinfo
  {volume} {78}},\ \bibinfo {pages} {184517} (\bibinfo {year}
  {2008})}\BibitemShut {NoStop}%
\end{thebibliography}%

\end{document}